\documentstyle[preprint,aps]{revtex}
\input epsf
\begin{document}
\draft
\title{Non-universal coarsening and universal distributions in
far-from equilibrium systems} 
\author{F. D. A. Aar\~ao Reis${}^{1,}$\footnote{Email address:
reis@if.uff.br} and R. B. Stinchcombe${}^{2,}$\footnote{E-mail address:
r.stinchcombe1@physics.ox.ac.uk}}
\address{
${}^{1}$ Instituto de F\'\i sica, Universidade Federal Fluminense,
Avenida Litor\^anea s/n, 24210-340 Niter\'oi RJ, Brazil\\
${}^{2}$ Theoretical Physics, Department of Physics, Oxford University, 1 Keble
Road, Oxford OX1 3NP}
\date{\today}
\maketitle
\begin{abstract}
Anomalous coarsening in far-from equilibrium one-dimensional
systems is investigated by
applying simulation and analytic techniques to minimal hard core particle
(exclusion) models. They contain mechanisms of aggregated particle diffusion,
with rates $\epsilon\ll 1$, particle deposition into cluster gaps, but
suppressed for the smallest gaps, and breakup of clusters which are adjacent
to large gaps. Cluster breakup rates vary with the cluster length $x$ as
$kx^\alpha$. The domain growth law $\langle x\rangle \sim
{\left( \epsilon t\right)}^z$, with $z=1/\left( 2+\alpha\right)$ for
$\alpha>0$, is explained by a simple scaling picture involving the
time for two particles to coalesce and a new particle be deposited.
The density of double vacancies, at which deposition and cluster breakup
are allowed, scale as $1/\left[ t{\left( \epsilon t\right)}^z\right]$.
Numerical simulations for several values of $\alpha$ and $\epsilon$
confirm these results. A fuller approach is presented which employs a
mapping of cluster configurations
to a column picture and an approximate factorization of the cluster
configuration probability within the resulting master equation. The
equation for a one-variable scaling function explains the above average
cluster length scaling. The probability distributions of cluster lengths $x$
scale as $P(x)=\frac{1}{{\left( \epsilon t\right)}^z}g(y)$,
with $y\equiv x/{\left( \epsilon t\right)}^z$, which is confirmed by
simulation. However, those distributions show a universal tail with the
form $g(y) \sim \exp{\left( -y^{3/2}\right)}$, which
is explained by the connection of the vacancy dynamics with the
problem of particle trapping in an infinite sea of traps. The high
correlations of surviving particle displacement in the latter problem
explains the failure of the independent cluster approximation to represent
those rare events.
\end{abstract}


\section{Introduction}
\label{introduction}

Domain growth in far from equilibrium systems is a subject of increasing
interest due to the large number of applications, such as phase separation of
mixtures, dynamics of glasses and island coarsening after thin film
deposition~\cite{ritort,mevansreview,robin}.  In these systems, their dynamics
is responsible for bringing them
to steady states, while external agents act to drive them out of equilibrium.
Many statistical models exhibit simple, universal domain growth laws, which are
found in some real systems, but there is much interest in models with slow
coarsening and with continuously varying growth exponents, for instance due to
their potential applications to glassy systems~\cite{ritort,mevansreview,dhar}.

In this paper, we will consider one-dimensional models with particle deposition
and diffusion, reversible aggregation to clusters and mechanisms of cluster
breaking which show such a variety of domain growth laws.  Cluster breaking may
be an effect of internal stress and was previously considered in studies of
island growth in submonolayers~\cite{koponen}.  In a real system subject to
external pressure but with some type of geometrical frustration other than those
observed in island growth, it is expected that cluster breaking will compete
with mechanisms of densification, these ones to be represented by vacancy
filling (deposition of new particles).  However, the onset of those processes
depends on the formation of large vacancies due to the (slow) diffusion of
aggregated particles.  The coarsening exponents of those systems will be shown
to vary with the exponents in the scaling of the cluster break probabilities,
although the cluster lengths distributions are universal.  Consequently, these
one-dimensional statistical models, although not related to a specific real
problem, reveal some interesting features that may help to understand complex
three-dimensional systems, with the advantage of being more tractable both
analytically and numerically.

The models presented here are nontrivial extensions of those analyzed in a recent
paper~\cite{coarsen1}, which include particle diffusion, reversible aggregation to
clusters and deposition mechanisms.  In the original model, hard-core particles in
a one-dimensional lattice have diffusion rates $r=d$ when they were free (i.  e.,
they have two empty nearest-neighbor sites) and $r=\epsilon\sim e^{-E/T}$ when they
have one occupied nearest neighbor site, with $r\ll d$ (Fig.  1a).  The deposition
rate is $F=1$, in units of monolayers per time step, and is restricted to sites
with at least one empty nearest neighbor (Fig.  1b), i.  e.  a site of a double or
larger vacancy.  These dynamical rules were motivated by the Clarke-Vvedensky model
and related models of thin films or submonolayer growth~\cite{cv}, but included
effects of geometrical frustration that forbid filling of single vacancies.  Domain
growth in the form $\langle d\rangle\sim \epsilon^{-1/2}t^{1/2}$ was predicted
analytically and from simulation~\cite{coarsen1}.  The same model without
deposition and in the limit $\epsilon\ll 1$ showed domain growth as $t^{1/3}$
before approaching a steady state~\cite{coarsen1}.

Here, in addition to the processes of the original model (Figs. 1a and 1b),
we will consider the competition between deposition of new particles and
the breaking of a neighboring cluster when a double (or larger) vacancy
appears. In this model, a
cluster of length $x$ with two vacancies at one of its sides (where deposition
may also occur) may break in two pieces, at a random internal position, with
rate $kx^{\alpha}$, where the exponent $\alpha$ is a tunable parameter and
$k$ is a constant amplitude. 
The process is illustrated in Fig. 1c for a cluster with $x=6$, with a
total of $5$ internal points for its separation into two pieces. Our focus is the
non-trivial case $\alpha >0$, for which $\alpha$-dependent coarsening exponents
are obtained.

Notice that this model is significantly different from other models which
involve coagulation or breakup of clusters with size-dependent rates with
variable exponents~\cite{family,meakin,leyvraz,bennaim}. Here, the clusters
slowly gain mass by (non-biased) diffusion at the expense of neighboring clusters,
while breakup is subject to the availability of free space for its expansion. 

Scaling laws for the average cluster length
$\langle x\rangle$ in the form
\begin{equation}
\langle x\rangle\sim t^z
\label{defz}
\end{equation}
will be obtained, where $z$ is called the coarsening exponent.  In the
models with the cluster breaking mechanism, the exponent $z$ can be
continuously tuned from $0$ to $1/2$ by varying the scaling exponent $\alpha$
of the rate of cluster breaking. This result is predicted by a scaling theory
that also described the cluster growth laws of Ref. \protect\cite{coarsen1}
and is confirmed by numerical simulations with very good accuracy. It is also
possible to describe such systems in terms of interval probabilities, which
allows the analytic investigation based on an independent cluster approximation.
Using this method, we also predict the coarsening exponents of the model.

There is also much interest in knowing the
distributions of cluster lengths in such anomalously growing systems. Those
distributions are also calculated numerically and show an universal
($\alpha$-independent)
form $P\left( x\right) \sim \exp{\left( -x^{3/2}\right)}$, despite the
fact that the coarsening exponents do depend on $\alpha$. The
analytical treatment of the model in the independent cluster approximation
gives instead an $\alpha$-dependent power in the exponent ($-x^{1+\alpha/2}$).
However, it
can be shown that the dynamics of large clusters is related to the problem
of particle diffusion in an infinite sea of mobile
traps~\cite{bramson,bray1,bray2}, where the holes between clusters in the
former problem correspond to the particles in the latter. This connection
leads to the above universal cluster length distribution and explains
the failure of the independent cluster approximation to predict those
distributions.

This paper is organized as follows. In Sec.  II we
present the scaling theory and obtain the coarsening exponents.
In Sec. III we present the results of simulations for the time dependence
of average cluster lengths and density of double vacancies. In Sec. IV,
we map the cluster configurations to a column picture and determine the
master equation with the independent interval approximation to the joint
cluster length probability. In Sec. V, we discuss the cluster length
probabilities, comparing numerical results, the analytical prediction of the
independent interval approximation and the connection to the problem of
one diffusing particle in a sea of moving traps. In Sec. VI, we summarize
our results and present our conclusions.

\section{Models and scaling theories}
\label{modelscaling}

Here we will define our models and estimate coarsening exponents using
scaling arguments along the lines of Ref. \protect{\cite{coarsen1}, which
were previously adopted in the analysis of related systems by
Evans~\cite{mevansreview} and introduced in the analysis of domain growth
in magnetic systems by Lai et al~\cite{lai} and Shore et al~\cite{shore}.

First we consider the original model of deposition and diffusion presented in
Ref. \protect{\cite{coarsen1} (Figs. 1a and 1b).

For simplicity, we will refer to
the average cluster length as $x$.  In Fig.  2a, we show a configuration with
clusters of lengths typically of order $x$, named $A$, $B$ and $C$, with single
empty sites (single vacancies) between them.  Deposition is not allowed at
those vacancies, as well as at the other single holes separating $A$ and $C$
from other neighboring clusters.  Suppose that the length of $A$ tends to
increase in time, while the length of $B$ decreases, as shown in Fig.  2b.
This evolution is equivalent to the diffusion of the vacancy between $A$ and
$B$, which gets closer to the vacancy between $B$ and $C$ (diffusion of this
vacancy was not shown only to simplify the illustration).  Finally, the
length of $A$ will increase by $x$ when $A$ and $B$ coalesce, as shown in Fig.
2c.  At that time, a new particle is deposited in an empty site of the double
vacancy, which is also shown in Fig.  2c.

The deposition process occurred after all particles of cluster $B$ have detached
from it and aggregated to cluster $A$.  This is equivalent to displacement of
order $x$ of the diffusing vacancy between $A$ and $B$.  Notice that in a
configuration with two neighboring empty sites (Fig.  2c), the probability of
deposition (fixed rate $F=1$) is much larger than the probability of diffusion
of an aggregated particle ($\epsilon \ll 1$), thus it is highly improbable that
a diffusion process will follow the formation of a double vacancy.
Consequently, the time necessary for coarsening of two clusters of length $x$
is of the order of the time for a diffusing vacancy to move a distance $x$,
which is $\Delta t \sim {\epsilon}^{-1}x^2$.  In the configuration after
deposition, the average cluster length is increased by $x$, since it was
equivalent to the merging of cluster $B$ into cluster $A$.  Consequently, the
average cluster length increases as
\begin{equation}
{{dx}\over{dt}} \sim {{\Delta x}\over{\Delta t}} \sim
{x\over{{\epsilon}^{-1}x^2}} = {\epsilon\over
x} .
\label{scalingor}
\end{equation}
Integrating Eq. \ref{scalingor}, we
obtain $x\sim {\epsilon}^{1/2} t^{1/2}$, in agreement with the analytical
results of the independent cluster approximation and simulation
data~\cite{coarsen1}.

Now consider the problem with the cluster breaking mechanism.
It is assumed that a cluster of length $x$ breaks with rate
\begin{equation}
r_{cb} = k x^{\alpha}
\label{ratecb}
\end{equation}
only
when there is enough space available, i. e., when there is more than one empty
site at one of the sides of the cluster. This process is illustrated in Fig. 1c,
in which a cluster of length $x=6$ may break at five different internal points,
which gives five possible final configurations if the breaking occurs -
three of them were shown in Fig. 1c.

Following the argument of the original model, after the diffusion processes
leading to the onset of a double vacancy (Fig. 2c), there are three
possibilities: a) a new particle is deposited with rate $F=1$; b) the cluster
at the right side of the double vacancy breaks; c) the cluster at the left side
breaks. Since the two neighboring clusters have lengths of order $x$,
deposition will occur with probability $P_{dep} \sim 1/x^{\alpha}$
for $\alpha>0$ and $x\ll 1$. Otherwise, one of the clusters adjacent to
the double vacancy will break in two pieces.

If a new particle is deposited, then the average
cluster length will increase by $x$, such as in model I.
On the other hand, in the
most probable case of an adjacent cluster to break, the average cluster
length will not increase. Consequently, the average cluster length
increases as
\begin{equation}
{{dx}\over{dt}} \sim P_{dep} {{\Delta x}\over{\Delta t}} \sim
\frac{1}{x^\alpha}
\frac{x}{{\epsilon}^{-1}x^2} = {\epsilon\over x^{1+\alpha }} .
\label{scalingcb}
\end{equation}
Integrating Eq. \ref{scalingcb}, we obtain
\begin{equation} x\sim {\epsilon t}^z , z = {1\over{2+\alpha }} .
\label{zcb}
\end{equation}

For $\alpha<0$, the above reasoning leads to $P_{dep}\approx 1$ for large $x$
and, consequently, $z=1/2$, as in the original model.

For $\alpha>0$, Eq.  \ref{zcb} shows that the coarsening exponent can be
continuously tuned from $z=0$ to $z=0.5$ by changing the cluster
breaking exponent.

The same arguments can be used to predict the density of double vacancies,
$\rho_{00}\left( t\right)$,
a quantity which plays a important role in the analytic calculations of
Sec. \ref{theory}. The total rate at which deposition and cluster-breaking
occur after the formation of a double vacancy (Fig. 2c) is of order
$x^\alpha$, for $\alpha>0$. Consequently, that vacancy will survive during
a time $1/x^\alpha$, while it takes a time of order 
$\Delta t \sim {\epsilon}^{-1}x^2$ to be formed (Figs. 2a-2c). Consequently,
one double vacancy between two clusters of length $x$ typically survives
during a fraction
${\left( 1/x^\alpha\right)}/{\left({\epsilon}^{-1}x^2\right)}$ of the
total time. To obtain its density we must divide this quantity by the
average cluster length $x$, which gives
\begin{equation}
\rho_{00}\left( t\right) \sim
\frac{\left( 1/x^\alpha\right)}{\left({\epsilon}^{-1}x^2\right)} \frac{1}{x}
\sim {\epsilon\over x^{3+\alpha}} \sim {{\epsilon t}^{-z}\over{t}} .
\label{scalingpoo}
\end{equation}
Notice that this time decay does not involve the diffusive factor
$\epsilon t$ alone.

\section{Simulation results for average lengths and densities}
\label{simulation}

We simulated the model for cluster-breaking exponents $\alpha=0.5$, 
$\alpha=1$, $\alpha=2$ and $\alpha=4$, with $\epsilon =0.001$ and, in some
cases, also with $\epsilon=0.01$. 
Deposition rates, free particle diffusion rates and amplitudes of
cluster breaking rates were $F=1$, $d=1$ and $k=1$, respectively, in all
simulations.
Lattices lengths were $L=5\times {10}^4$.
The maximum simulation times were ${10}^6$ for the smaller value of $\alpha$
and ${10}^8$ for the largest one. Thus, at all times, the
whole lattice contained more than $500$ clusters, which was essential to
avoid finite-size effects. Indeed, several results were compared with those
in lattices of lengths $L={10}^4$ and confirmed the absence of
significant finite-size effects. The number of different realizations
varied from $5000$ for the smaller $\alpha$ to $500$ for the largest one.

The scaling of the average cluster length $\langle x\rangle$ on $\epsilon$
and $t$ is confirmed in Fig. 3 for $\alpha=1$, by showing
$\langle x\rangle/{\left( \epsilon t\right)}^{1/3}$ for $\epsilon =0.01$
and $\epsilon =0.001$ as a function of
$1/{\left( \epsilon t\right)}^{1/3}$. The data collapse confirms the expected
scaling on $\epsilon$, the convergence to a finite value as $t\to\infty$ 
confirms the scaling on $t$ and the variable in the abscissa suggests
a constant correction term to Eq. (\ref{scalingcb}).

In Fig. 4a and 4b we show $\langle x\rangle/{\left( \epsilon t\right)}^z$ 
versus $1/{\left( \epsilon t\right)}^z$ for
$\alpha=0.5$ and $\alpha =2$, respectively, with the values of $z$ given
by Eq. (\ref{scalingcb}) and $\epsilon = 0.001$. These results confirm
the validity of the scaling theory of Sec. \ref{modelscaling} for the
average cluster length.

Since double vacancies are very rare, the accuracy of their density is much
lower than the accuracy of $\langle x\rangle$. Consequently, averages over
relatively large time regions were necessary to study the long time evolution
of that density. The average values agree with 
the predicted scaling (Eq. \ref{scalingpoo}),
as shown in Fig. 5 for $\alpha=1$ (with two values of $\epsilon$).
Notice that the different time and $\epsilon$ dependences were confirmed there.

\section{Theoretical analysis under the independent cluster approximation}
\label{theory}

Now we turn next to the more powerful analysis starting from a version of the
master equation, which can provide a full description of the process. This is
more easily set up by reformulating the process using a column picture, in
which a column of height $m$ represents a cluster of size $m$. The
map of the problem of particles in a line onto the cluster problem in a line
of reduced length is shown in Figs. 6a and 6b. The
original diffusion processes of free and aggregated particles correspond
to those shown in Fig. 6b. Since one cluster has two edges but
corresponds to a single column, the
one-particle detachment rate in the column picture is
\begin{equation}
\gamma = 2\epsilon .
\label{rates}
\end{equation}
Finally, in Fig. 7 we illustrate the competition between deposition of a
new particle and cluster breaking after the formation of a double vacancy
in the particle picture, which corresponds to the formation of a single
vacancy in the cluster picture.

The deposition process (Fig. 7) leads to the decrease of the total number of
clusters and of the number of holes between the clusters as time increases.
On the other hand, the length $L$ of the line in which particles are
deposited and diffuse is kept constant. Consequently, in order to adopt
the column picture, it is necessary to consider that
the length $L_0$ of the corresponding column problem decreases in time.
These lengths are related as $L_0=L-M$, where $M$ is the total mass or
total number of particles, for periodic boundary conditions.

We denote by $P_t(m)$ the probability that a randomly chosen cluster
(equivalently, column) has size $m$ at time $t$. It is given in terms of
the clusters numbers $N(m,t)$ by
\begin{equation}
P_t(m) = {{N(m,t)}\over{L_0(t)}} ,
\label{probdifdep}
\end{equation}
while the length of the lattice in which the column problem is defined
varies due to deposition as
\begin{equation}
{{L_0(t+1)-L_0(t)}\over{L_0(t)}} = -P_t(0) \left[ 2-P_t(0)\right] .
\label{lenght}
\end{equation}
Here, $P_t(0)$ is the probability of an empty site in the column problem,
which correspond to double vacancies in the original particle problem (see
Figs. 6 and 7).

Then the gain/loss from in/out processes provides a master equation which
must be written in terms of clusters numbers $N(m,t)$ in the general form
\begin{equation}
N(m,t+1) - N(m,t) = L_0 \left( B_m + C_m + D_m + {\cal B}_m \right) ,
\label{masterequation}
\end{equation}
where $B_m$ comes from the cluster breaking processes, $C_m$ comes from the
diffusion of aggregated particles (detachment of particles from clusters),
$D_m$ comes from the diffusion of free particles and ${\cal B}_m$ comes from
the deposition processes. The terms in the right-hand side of Eq.
(\ref{masterequation}) are thus written in terms of probabilities of cluster
masses at time $t$.

In an independent
interval approximation in which joint probabilities are factorized, they
are given by
\begin{eqnarray}
C_m &=& \gamma \theta (m+1-2) P_t(m+1) +
\gamma \theta (m-1) P_t(m-1)\left[ 1-P_t(0)-P_t(1)\right] \nonumber\\
&& -\gamma \theta (m-2) P_t(m) - \gamma P_t(m)\left[ 1-P_t(0)-P_t(1)\right] ,
\label{masterdifag}
\end{eqnarray}
\begin{eqnarray}
D_m &=& d \theta (m-1) P_t(m-1) P_t(1) + d\delta_{m,0}P_t(1) -
d\delta_{m,1}P_t(1) - d P_t(m) P_t(1)\nonumber\\
&=& dP_t(1) \left[ \theta (m-1) P_t(m-1) + \delta_{m,0} -
\delta_{m,1} - P_t(m) \right] ,
\label{masterdiflivre}
\end{eqnarray}
\begin{equation}
{\cal B}_m = P_t(0) \left[ 2 \theta\left( m-2\right) P_t\left( m-1\right) -
2 \theta\left( m-1\right) P_t\left( m\right) + \delta_{m,1}P_t\left( 0\right)
- 2\delta_{m,0} \right] 
\label{masterdep}
\end{equation}
and
\begin{eqnarray}
B_m &=& 2 \theta (m-1) P_t(0) \sum_{r=1}^{\infty}{ f(m+r) P_t(m+r)} \nonumber\\
&&- \theta (m-2) f(m) P_t(m) (m-1) P_t(0)
 -\delta_{m,0} P_t(0) \sum_{p=2}^{\infty}{ (p-1) f(p) P_t(p)} \nonumber\\
&=& P_t(0) [
2\theta (m-1) \sum_{r=1}^{\infty}{ f(m+r) P_t(m+r)}
\nonumber\\
&& - \theta (m-2) (m-1) f(m) P_t(m)
- \delta_{m,0} \sum_{p=2}^{\infty}{ (p-1) f(p) P_t(p)}
] .
\label{mastercb}
\end{eqnarray}
In Eq. (\ref{mastercb}), the function $f(m)$ gives the rate for a cluster of
mass $m$ to break at each internal point, i. e. at each internal connection
between two aggregated particles:
\begin{equation}
f(m) = k \frac{{\left( m-1\right)}^\alpha}{m-1} =
k {\left( m-1\right)}^{\alpha -1} .
\label{fm}
\end{equation}

The generating function
\begin{equation}
G_t(s) \equiv \sum_{m=0}^{\infty}{P_t(m) s^m}
\label{generating}
\end{equation}
satisfies
\begin{equation}
G_{t+1}(s) {\left[ 1-P_t(0)\right]}^2 - G_t(s) =
{\cal L}\left( B + C + D + {\cal B}\right) ,
\label{difgenerating}
\end{equation}
where the operator ${\cal L}$ is defined as
\begin{equation}
{\cal L}g = \sum_{m=0}^{\infty}{g(m) s^m} .
\label{laplace}
\end{equation}
We thus obtain
\begin{equation}
{\cal L} C = {\gamma\over s}\left( 1-s\right) \left[ G_t(s) - P_t(0)
- P_t(1) s \right] + \left( s-1\right) \gamma b G_t(s) ,
\label{laplacedifag}
\end{equation}
with
\begin{equation}
b\equiv 1-P_t(0)-P_t(1) ,
\label{defb}
\end{equation}
\begin{equation}
{\cal L} D = d P_t(1) \left( s-1\right) \left[ G_t(s)-1\right] ,
\label{laplacediflivre}
\end{equation}
\begin{equation}
{\cal L} {\cal B} = P_t(0) \left[ 2\left( s-1\right)
\left( G_t(s)-P_t(0) \right) + sP_t(0) - 2 \right]
\label{laplacedep}
\end{equation}
and
\begin{eqnarray}
{\cal L} B &=& 2P_t(0) \sum_{m=1}^{\infty}{ \left[ s^m
\sum_{r=0}^{\infty}{ f(m+r) P_t(m+r) } \right] } -
P_t(0) \sum_{m=2}^{\infty}{ s^m \left( m-1\right) f(m) P_t(m)} \nonumber\\
&& - P_t(0) \sum_{m=2}^{\infty}{ \left( p-1\right) f(p) P_t(p)} .
\label{laplacecb}
\end{eqnarray}
Eqs. (\ref{laplacedifag}), (\ref{laplacediflivre}) and (\ref{laplacedep})
are the same obtained in Ref. \protect\cite{coarsen1} for the model
with only deposition and diffusion (with $d=1$).

Now defining the operator ${\cal F}$ so that
\begin{equation}
{\cal F}\left( s{\partial\over{\partial s}} \right) \sum_m{ e^{sm} g(m)} =
\sum_m{ f(m) e^{sm} g(m)} ,
\label{depopf}
\end{equation}
we may write the contribution of cluster breaking processes to the generating
function as
\begin{eqnarray}
{\cal L} B &=& P_t(0) \frac{\left( 1+s\right)}{\left( 1-s\right)}
\left[
{\left[ {\cal F}\left( s{\partial\over{\partial s}} \right) G_t(s)
\right] }_{s=1} -
{\cal F}\left( s{\partial\over{\partial s}} \right) G_t(s)
\right]
\nonumber\\
&& - P_t(0) \left[
{\left[
s{\partial\over{\partial s}}
{\cal F} \left( s{\partial\over{\partial s}} \right) G_t(s)
\right]}_{s=1}
+ s{\partial\over{\partial s}}
{\cal F}\left( s{\partial\over{\partial s}} \right) G_t(s)
\right].
\label{laplacecb1}
\end{eqnarray}

Because deposition slowly fills the system, we expect the configurations to
coarsen and presumably to go into some scaling asymptotics where mass scales
with some power of $t$, and $P_t(m)$ and $G_t(s)$ each become one-variable
scaling functions. So we look for a long time scaling solution of the above
equation, in which the finite difference $G_{t+1}(s)-G_t(s)$ in Eq.
(\ref{difgenerating}) can be taken as a time derivative. The scaling variable
will be
some combination of $t$ (large) and $u\equiv 1-s$ (small), the latter because
large cluster sizes arise from structure in $G_t(s)$ at $s\approx 1$. The
variable $u$ is actually conjugate to $m$. Coarsening will
correspond to the scale of $m$ as $t^z$, with some power $z$, in which case
the one-variable form will be
\begin{equation}
G_t(s) = u^\alpha F\left( ut^z\right) ,
\label{scalinggenerating}
\end{equation}
with some function $F$.
Normalization requires $\alpha = 0$ and $F(0) = 1$. In the scaling limit, the
relationship of the generating function to the probability $P_t(m)$ requires
the latter to be of the form
\begin{equation}
P_t(m) = {1\over{t^z}} g\left( {m\over {t^z}}\right) ,
\label{scalingprob}
\end{equation}
with
\begin{equation}
F(x) = \int_0^\infty{ g(y) e^{-xy} dy} .
\label{deff}
\end{equation}
Moreover, for $s\approx 1$ one has $s{\partial\over{\partial s}} \approx
{\partial\over{\partial u}}$ in Eq. (\ref{laplacecb1}).

Thus we obtain the following equation for the one-variable scaling function:
\begin{eqnarray}
\lefteqn{
\left[ {\left( 1-\frac{1}{t^z}g(0) \right)}^2
\left( 1+{\partial\over{\partial t}}\right) -1\right] F{\left( ut^z\right)} =}
\nonumber\\
&&\begin{array}{r}
u \left[ F{\left( ut^z\right)} \left( a-\gamma \left( 1-u+u^2+\dots \right)
\right) +
\left( \gamma -1\right) \frac{1}{t^z} g{\left( \frac{1}{t^z}\right)}
+ \gamma \left( 1-u+u^2+\dots \right) \frac{1}{t^z} g(0) \right]
\\
+ \frac{1}{t^z} g(0) \left[ 2u F{\left( ut^z\right)} -
\left( 1-u \right) \frac{1}{t^z} g(0) + \frac{2}{t^z} g(0) -2 \right]
\\
+ \frac{1}{t^z} g(0) \left[ -\frac{\left( 2+u\right)}{u}
{\left[ {\cal F}\left( {\partial\over{\partial u}} \right) F{\left( ut^z\right)}
\right]}_{u=0} -
{\left[ {\partial\over{\partial u}}
{\cal F}\left( {\partial\over{\partial u}} \right)
F{\left( ut^z\right)} \right]}_{u=0} \right]
\\
- \frac{1}{t^z} g(0) 
\left( {\partial\over{\partial u}} - \frac{\left( 2+u\right)}{u} \right)
F{\left( ut^z\right)} ,
\end{array}
\label{eqF}
\end{eqnarray}
where $a\equiv \gamma - \gamma P_t(0) + \left( 1-\gamma\right) P_t(1)$.

The dominant terms in Eq. \ref{eqF} give
\begin{eqnarray}
\frac{zx}{t} F'(x) - 2P_t(0) F(x) &=& \gamma {\left( \frac{x}{t^z}\right)}^2
F(x) - 2P_t(0)
\nonumber\\
&&
+ P_t(0) t^z \left[ -\frac{2}{x} {\left[ 
{\cal F}\left( t^z \frac{d}{dx} \right) F(x) \right]}_{x=0}
- {\left[ \frac{d}{dx}
{\cal F}\left( t^z \frac{d}{dx} \right) F(x) \right]}_{x=0} \right]
\nonumber\\
&&+ P_t(0) t^z \left( \frac{2}{x}- \frac{d}{dx} \right)
{\cal F}\left( t^z \frac{d}{dx} \right) F(x) .
\label{eqFdominant}
\end{eqnarray}

In the scaling limit, the first term at the left hand side (LHS) is of order
$1/t$. Assuming that the density of clusters of zero mass (holes in the column
picture) decays as
\begin{equation}
P_t(0) \sim \frac{1}{t^\beta} ,
\label{defbeta}
\end{equation}
the second term at the LHS and the second term at the right hand side
(RHS) of Eq. (\ref{eqFdominant}) are both of order $1/t^\beta$.
The first term in the RHS, related to the
diffusion of aggregated particles, is of order $1/t^{2z}$.
Since cluster breaking competes with deposition and recalling that the
coarsening exponent in the problem with only deposition and diffusion is $z=1/2$,
we expect that, in the present system, $z\leq 1/2$ for any $\alpha$.
The last term in the RHS, associated with cluster breaking, is of order
$1/t^{\left( \beta -z\alpha\right)}$, which clearly dominates over the terms
which scale as $1/t^\beta$ for any $\alpha >0$.
Thus, we conclude that the terms associated with aggregated particles diffusion
(first of the RHS of Eq. \ref{eqFdominant}) and with cluster breaking
(last of the RHS) are dominant for $\alpha >0$ and must cancel each other,
while the remaining
terms are subdominant contribution which must also cancel. From the dominant
terms, we obtain
\begin{equation}
2z = \beta - z\alpha
\label{expdom}
\end{equation}
and from the subdominant terms we obtain
\begin{equation}
\beta = 1 ,
\label{expsubdom}
\end{equation}
which gives the coarsening exponent
\begin{equation}
z = 1 / \left( 2+\alpha \right) .
\label{zteorico}
\end{equation}

The value of $z$ agrees with that obtained from scaling arguments and in
simulations. In order to compare the density of single vacancies in the column
problem, $P_t(0)$, and the density of double vacancies in the original
particle problem, $\rho_{00}\left( t\right)$, we consider that the total length
of the lattice in the column problem, $L_0$, is smaller than the length $L$
by a factor of the order of the average cluster size (see Figs. 6 and 7),
which scales as $t^z$.
Consequently
\begin{equation}
\rho_{00}\left( t\right) \sim P_t(0) / t^z \sim t^{-\left( 1+z\right)} ,
\label{reldensities}
\end{equation}
which also agrees with the scaling arguments of Sec. II and simulation results.
The $\epsilon$- or $\gamma$-dependence of average cluster size and densities
of vacancies in both pictures, predicted in Secs. II and III, also follow from 
Eq. (\ref{eqFdominant}).

Also notice that, for $\alpha<0$, the cluster breaking term of 
Eq. (\ref{eqFdominant}) is not dominant anymore and we obtain $z=1/2$, as in
the problem without that mechanism (Ref. \protect\cite{coarsen1}).

\section{Distributions of cluster lengths}
\label{distributions}

An equation for the one-variable scaling function $F(x)$ follows from the
balance of the dominant terms in Eq. (\ref{eqFdominant}), related to diffusion
of aggregated particles and cluster breaking:
\begin{eqnarray}
\frac{\gamma x^2 F(x)}{t^{2z}} &=&
P_t(0) t^z \left[ \frac{2}{x} {\left[ 
{\cal F}\left( -t^z \frac{d}{dx} \right) F(x) \right]}_{x=0}
+ {\left[ \frac{d}{dx}
{\cal F}\left( -t^z \frac{d}{dx} \right) F(x) \right]}_{x=0} \right]
\nonumber\\
&&+ P_t(0) t^z \left( \frac{d}{dx} - \frac{2}{x} \right)
{\cal F}\left( -t^z \frac{d}{dx} \right) F(x)
\label{eqFdominant1}
\end{eqnarray}
Physically, it means that the dynamics in
long time regime is dominated by diffusion of aggregated particles,
which forms larger clusters at the expenses of neighboring ones, and cluster
breaking, which restores a relatively random size distribution after the onset
of a vacancy in the column picture (double vacancy in the particle picture).
Deposition becomes a rare process, as expected with the large probabilities of
cluster breaking for $\alpha >0$, and does not contribute to determine the
main features of the scaling function.

Instead of solving Eq. (\ref{eqFdominant1}) for the function $F(x)$, which
would involve fractional derivatives for non-integer exponents $\alpha$, we
consider the same balance of contributions of diffusion of aggregated particles
and cluster breaking in the original master equation (\ref{masterequation}),
(\ref{masterdifag}) and (\ref{mastercb}). They
can be rewritten in terms of the function $g(y)$ defined in Eq.
(\ref{scalingprob}). At this point, we must consider that $P_t(0)$ scales as
\begin{equation}
P_t(0) \approx \frac{\left( k_0/k \right)}{t} ,
\label{scalingpo}
\end{equation}
where $k_0 \sim {\it o}(1)$, since the survival time of those vacancies 
is inversely proportional to the amplitude $k$ of the rate of cluster breaking
(Eq. \ref{ratecb}). We are thus led to
\begin{eqnarray}
\frac{\gamma}{t^{2z}} \frac{d^2g(y)}{dy^2}
+ \frac{\left( k_0/k\right)}{t} t^z
kt^{z\left( \alpha -1\right)}
\left[
2\int_y^\infty{dy' {y'}^{\left( \alpha-1\right)} g(y')}
- y^\alpha g(y) - \delta (y)
\int_0^\infty{dy' {y'}^{\left( \alpha-1\right)} g(y')} \right]
\nonumber\\
= 0 .
\label{eqg}
\end{eqnarray}

For $y>0$, Eq. (\ref{eqg}) leads to
\begin{equation}
\gamma_1 g''' = y^\alpha g' + \left( 2+\alpha \right) y^{\alpha-1} g ,
\label{eqdifg}
\end{equation}
with $\gamma_1=\gamma /k_0$.
From this equation we obtain the distribution in the form
\begin{equation}
g(y) \sim y^\mu \exp{\left( -Ay^\nu \right) },
\label{disttheory}
\end{equation}
with
\begin{equation}
\nu = 1+\alpha/2 , \mu = 1-\alpha/4 , A \sim {\gamma}^{-1/2} ,
\label{exptheory}
\end{equation}
which suggests that the shapes of the distributions of cluster lengths also
depend on the cluster breaking exponents.

This prediction can be compared with simulation results.
In all cases, the distributions were obtained by taking the length of
each cluster only once while spanning the lattice at a fixed simulation
time. Thus our estimates can be directly compared with the predictions from
the column picture. On the other hand, if a site average had been done during 
the simulations, a rescaling of the predictions of the column picture would be
necessary.

In Figs. 8a and 8b we show the scaled
cluster length distributions for $\alpha=2$ and $\alpha=4$, respectively.
From Eqs. (\ref{disttheory}) and (\ref{exptheory}), it is expected that
$\ln{\left[ {\left( \epsilon t\right)}^z P_t(m)/y^\mu \right]}$ decreases
linearly with $y^\nu$, with
\begin{equation}
y \equiv m/{\left( \epsilon t\right)}^z
\label{defy}
\end{equation}
($\mu=1/2$ and $\nu=2$ for $\alpha=2$, $\mu=0$ and $\nu=3$ for $\alpha=4$).
Here a factor $\epsilon^{-z}$ multiplies the variable in Eq.
(\ref{scalingprob}).

However, the curved shapes of the plots in Figs. 8a and 8b
disagree with the predictions of this independent cluster approximation.
On the other hand, it was observed that the data for all values of $\alpha$ can
be fitted by a universal distribution of the form
\begin{equation}
g(y) \equiv {\left( \epsilon t\right)}^z P_t(m) \sim
\exp{\left( -y^{3/2}\right) } .
\label{univdist}
\end{equation}
This is illustrated in Figs. 9a and 9b for $\alpha=0.5$ and $\alpha=1$, 
respectively, in
which the deviations from Eqs. (\ref{disttheory}) and (\ref{exptheory}) are
relatively small (indeed, for $\alpha=1$, the exponent $3/2$ of the universal
distribution is also predicted by Eq. \ref{exptheory}), and in Figs. 10a and 10b
for $\alpha=2$ and $\alpha=4$, respectively. Least squares fits of the data
for the largest lengths (dashed lines shifted two units to the right in
Figs. 9 and 10) confirm the validity of this universal scaling.

In order to explain this unexpected result, we must pay attention to
the details of the
dynamical process during the coarsening process, which comes from a balance
between aggregated particle diffusion and cluster breaking, and its connection
to diffusion-reaction related problems. Turning back to the original particle
picture of the problem, during almost all the time, the
dynamics is equivalent to diffusion of vacancies and scattering of one vacancy
upon collision of two or more vacancies. This scattering of a vacancy corresponds to
the breakup of a cluster at a random internal point (see Figs. 2 and 7).
Thus, while diffusion of a vacancy is the mechanism which leads to the formation
of large clusters locally, the collision of vacancies restores random
distribution of cluster lengths around the average value.

Focusing on a specific diffusing vacancy, it may be viewed as a particle in
an infinite sea of traps~\cite{bramson,bray1,bray2}, the latter
represented by the other vacancies. Subsequent
collisions approximately correspond to the random walk starting in the
origin and the absorption by a trap after a certain time - however, contrary to
the trapping process, the vacancy is thrown away and not absorbed. Thus the
probability of a cluster of length $x$ in our model can be approximated
by the probability of finding a particle at distance $x$ from the origin in
the trapping problem with an infinite sea of mobile traps (TP).

In Refs. \protect\cite{bray1} and \protect\cite{bray2},
the TP was studied analytically and it was
shown that the probability of the particle surviving at time $t$, being
always located in the interval $\left[ -l/2,l/2\right]$ and that
no trap has entered this region up to time $t$ is given by
\begin{equation}
P_{TP}(t) \sim \exp{\left[ -\lambda_1\rho_T {\epsilon t}^{1/2} -
\left( \rho_T l + \lambda_2 {\epsilon t}/l^2 \right) \right]} .
\label{probtrap}
\end{equation} 
In Eq. (\ref{probtrap}), $\rho_T$ is the
density of traps (of order of the inverse average cluster length,
$\langle x\rangle$ in our problem), $\lambda_1$ and $\lambda_2$ are
constants of order of a unit and it is assumed that the diffusion
coefficients of the particles and the traps are both equal to $\epsilon$.
This result was used in Ref. \protect\cite{bray1} to estimate a lower
bound for the probability of survival of a particle at time $t$ and it
was shown to converge to the upper bound they also estimated, thus providing
an exact limit for that probability at long times.

For a given time $t$, that distribution is peaked around $l=l_M$, where
\begin{equation}
l_M = {\left[ \frac{\lambda_2 \epsilon t}{\rho_T} \right]}^{1/3} .
\label{lmax}
\end{equation}
Instead of the broad displacement distribution of a free
particle problem, the surviving particles at time $t$ are much more
probably located
around the distance $l_M$ from the origin. Thus, the probability of
finding a surviving particle, which decays in time, may be expressed in
terms of this most probable distance $l_M$ from the origin.
In our problem of a diffusing vacancy, it corresponds to finding a cluster of
length $x\sim l_M$ at that time after the last collision with another vacancy.
This gives a distribution for cluster lengths
\begin{equation}
P{\left( x\right)}
\sim \exp{\left[ -\frac{\lambda_1}{{\lambda_2}^{1/2}} {\rho_T}^{3/2}
x^{3/2} - 3 \rho_T x \right]} \sim \exp{\left[ -C{\left( x/\langle x\rangle
\right)}^{3/2} \right]} .
\label{disttrap}
\end{equation} 
where we considered that $\rho_T\sim 1/\langle x\rangle$ and $C$ is a constant.

The above result agrees with those from simulation for all values of the
parameter $\alpha$. In its derivation,
a time average was performed, which corresponds to a time
average in the cluster breaking problem within a large time range for single
vacancy diffusion, but small for deposition of new particles and coarsening.
Indeed, this is the suitable interpretation to the fixed-time results of
Figs. 8-10, which were obtained from several snapshots of the system
near the times $t$ given above the plots. 

The discrepancy from the prediction of the theoretical
analysis based on the independent interval approximation is related to the
particular form of the distribution of surviving particles displacements in
the TP. The former approximation does not capture the correlated nature of the
rare processes.

Finally, we observe that the data for several values of $\alpha$ in Figs. 9 and
10, although lying approximately in the same region, do not collapse into a
single curve, which indicates the presence of corrections to the scaling
in Eq. (\ref{univdist}) which do depend on $\alpha$.

\section{Summary and conclusion}

We studied one-dimensional exclusion models with
particle diffusion, reversible attachment to clusters and
deposition mechanisms at large vacancies competing with breakup of
neighboring clusters. These models aim at representing coarsening of
aggregates subject to internal stress, so that the increase of density
when there is free space available is restricted by breakup that leads to
internal relief.

Different dependencies of the cluster breaking rate on the cluster length
were considered, involving the cluster breaking exponent $\alpha$.
Simulation results show cluster size growth as
$x\sim t^{1/\left( 2+\alpha\right)}$, which was explained using heuristic
scaling arguments. The analytical treatment of the master equation with an
independent cluster approximation for joint probabilities distributions
supports this prediction, as well as the scaling of the density of
double vacancies obtained numerically.

Despite the dependence of the coarsening exponent on the exponent $\alpha$,
universal probability distributions of large clusters were obtained,
in the form $P(x)\sim \exp{\left( -x^{3/2}\right)}$. This
result shows limitations of the independent interval approximation for the
treatment of rare events (cluster lengths much larger than the average value)
which emerge from the diffusion of vacancies and their survival at long
times without being scattered by the collision with other vacancies,
which corresponds to cluster breaking and restores a random distribution
of cluster lengths. However, the connection of the problem with the problem
of a particle diffusing in an infinite sea of mobile particles provides an
explanation for that distribution, which may be viewed as a snapshot of
the system at time intervals during which the coarsening process was
negligible.

We expect that the models presented above and the combination of different
methods to explain their scaling behaviors can be used to understand further
non-equilibrium systems. A particularly interesting application would be
the study of two-dimensional systems subject to the same conditions, in
order to describe coarsening of adatom islands in cases where there is a
mismatch of lattice parameters with the substrate and consequent stress or
strain of those islands.

\acknowledgements

FDAA Reis thanks the Department of Theoretical Physics at Oxford University,
where part of this work was done, for the hospitality, and acknowledges
support by CNPq and FAPERJ (Brazilian agencies).

RB Stinchcombe acknowledges support
from the EPSRC under the Oxford Condensed Matter Theory Grants,
numbers GR/R83712/01 and GR/M04426.

\begin{figure}
\epsfxsize=12cm
\begin{center}
\leavevmode
\epsffile{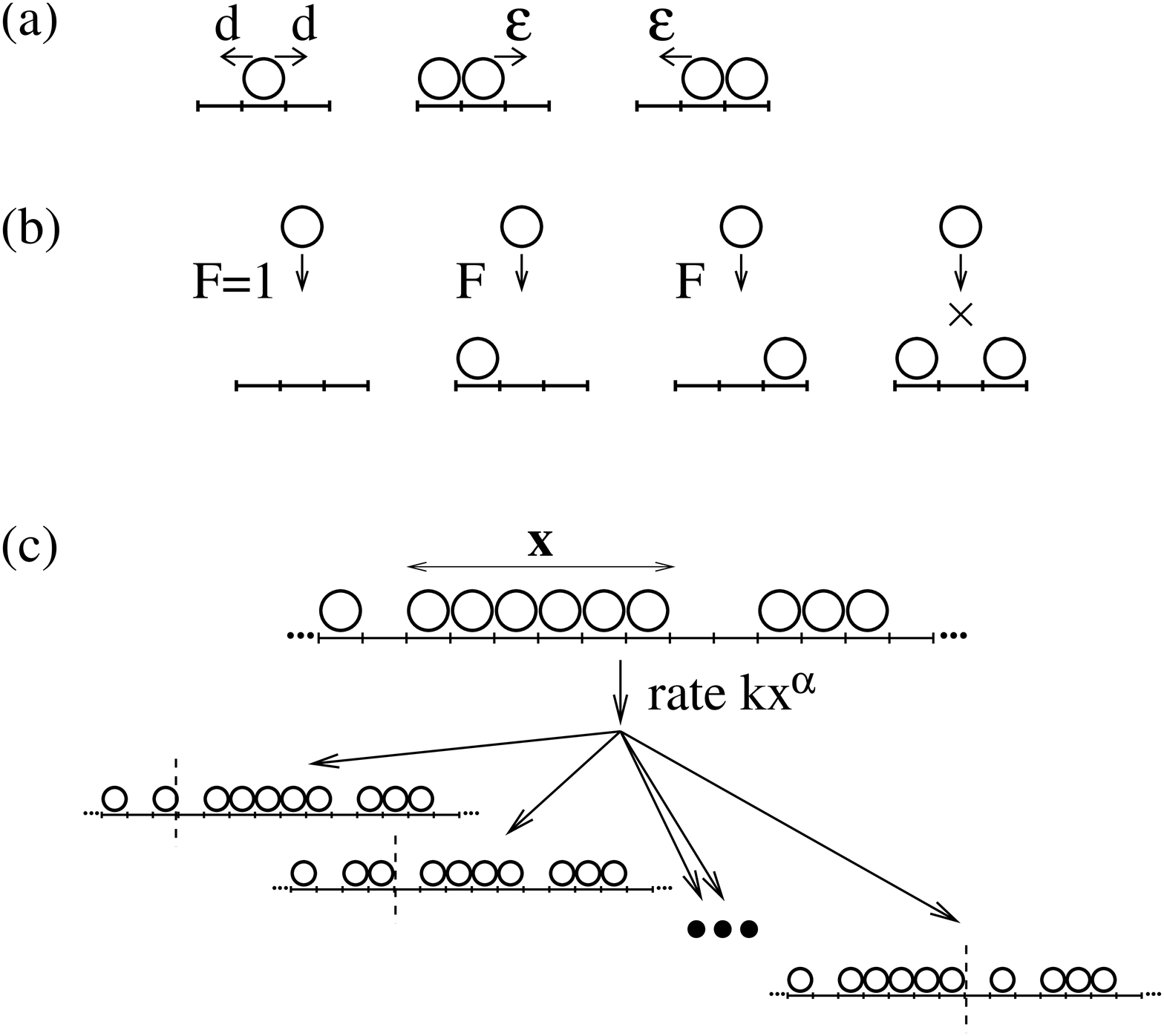}
\caption{(a) Diffusion processes of free and aggregated
particles at the borders of clusters, with the respective diffusion
rates. (b) Allowed deposition processes at vacancies with a neighboring vacant
site, with the respective rate, and the forbidden deposition process, in which
the vacancy has two occupied neighbors. (c) Three of the possible five
configurations after the breakup of the cluster at the left side of the
double vacancy, with the respective total rate. The breaking positions are
indicated by a dashed lines.}
\label{fig1}                        
\end{center}
\end{figure}     

\begin{figure}
\epsfxsize=12cm
\begin{center}
\leavevmode
\epsffile{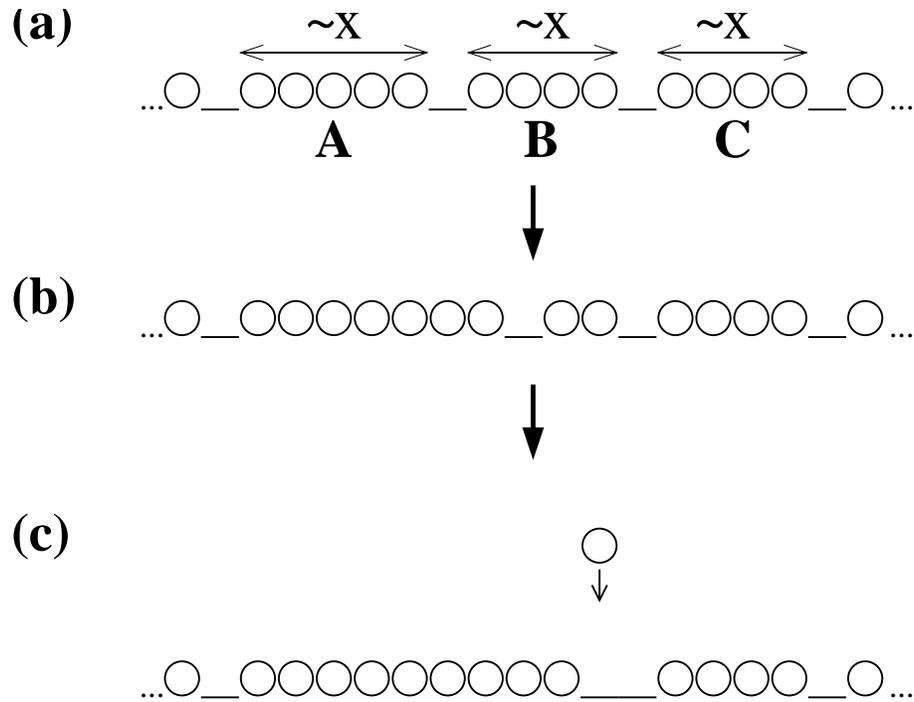}
\caption{(a) Three typical neighboring clusters, with lengths of the order
of the average cluster size $x$. (b) Growth of cluster A at the expenses of
cluster B, which is equivalent to the diffusion of the vacancy between them.
For simplicity, the effective diffusion of the vacancy between clusters B and C
was not illustrated. (c) Possible deposition after collision of the two
vacancies, extinction of cluster B and coarsening of cluster A.}
\label{fig2}                        
\end{center}
\end{figure}

\begin{figure}
\epsfxsize=12cm
\begin{center}
\leavevmode
\epsffile{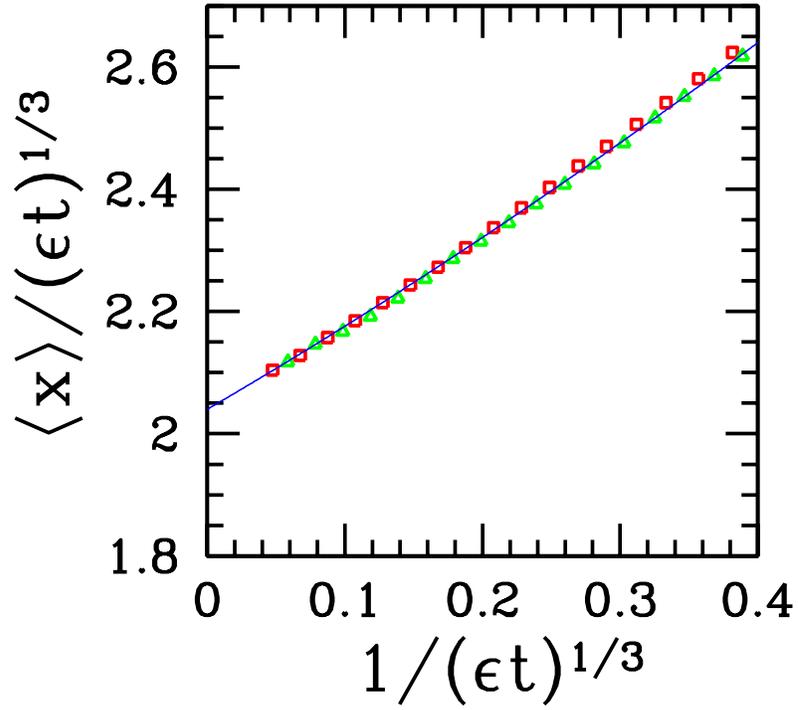}
\caption{Scaled cluster size as a function of scaled time for the model
with $\alpha=1$ and aggregated particle diffusion rates $\epsilon=0.001$
(squares) and $\epsilon =0.01$ (triangles). The solid curve is a parabolic fit
of the data for $\epsilon=0.001$.}
\label{fig3}                        
\end{center}
\end{figure}

\begin{figure}
\epsfxsize=12cm
\begin{center}
\leavevmode
\epsffile{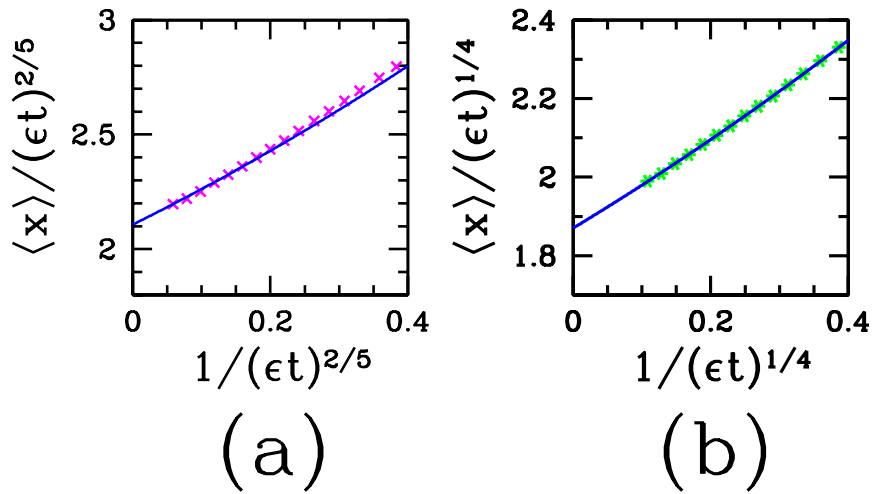}
\caption{Scaled cluster size as a function of scaled time for the model
with: (a) $\alpha=0.5$ and $\epsilon=0.001$; (b) $\alpha=2$ and
$\epsilon=0.001$. The solid curves are parabolic fits of the data.}
\label{fig4}                        
\end{center}
\end{figure}

\begin{figure}
\epsfxsize=12cm
\begin{center}
\leavevmode
\epsffile{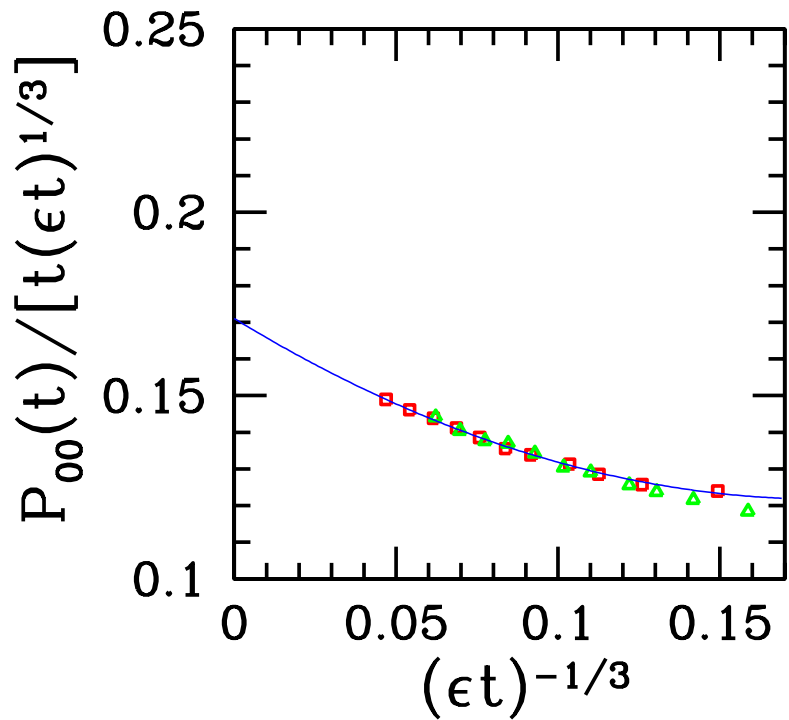}
\caption{Scaled probability of double vacancies as a function of scaled
time for the model
with $\alpha=1$ and aggregated particle diffusion rates $\epsilon=0.001$
(squares) and $\epsilon =0.01$ (triangles). The solid curve is a parabolic fit
of the data for $\epsilon=0.001$.}
\label{fig5}                        
\end{center}
\end{figure}

\begin{figure}
\epsfxsize=12cm
\begin{center}
\leavevmode
\epsffile{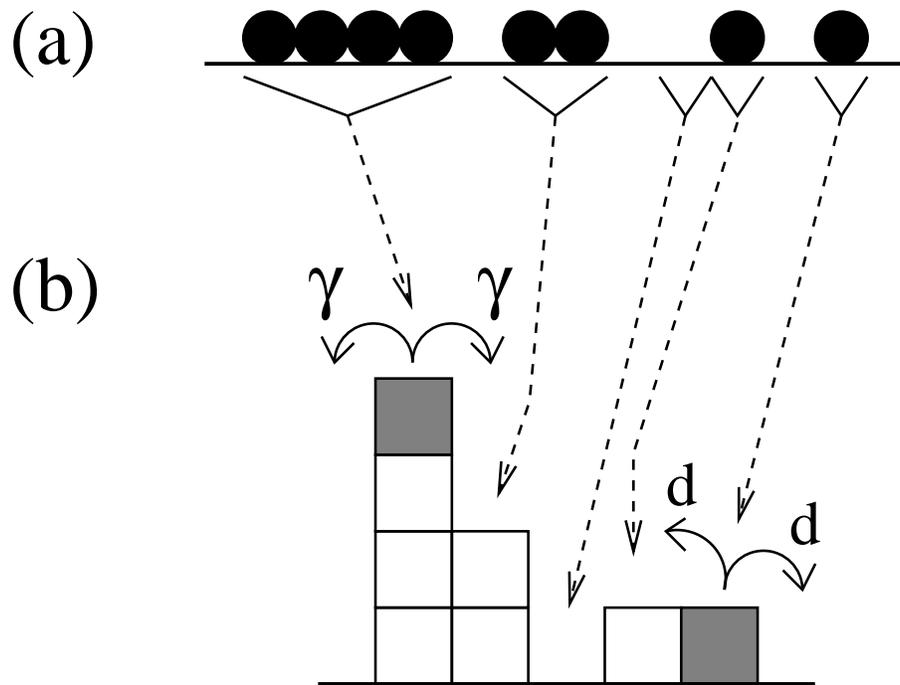}
\caption{(a) Example of particle-hole configuration on a line and the map
(dashed arrows) into a column problem. (b) The processes of particle
detachment from clusters, with rate $\gamma$, and of free particle
diffusion, with rate $1$, in the corresponding column picture. Filled squares
are the particles whose diffusion rates were indicated.}
\label{fig6}                        
\end{center}
\end{figure}

\begin{figure}
\epsfxsize=12cm
\begin{center}
\leavevmode
\epsffile{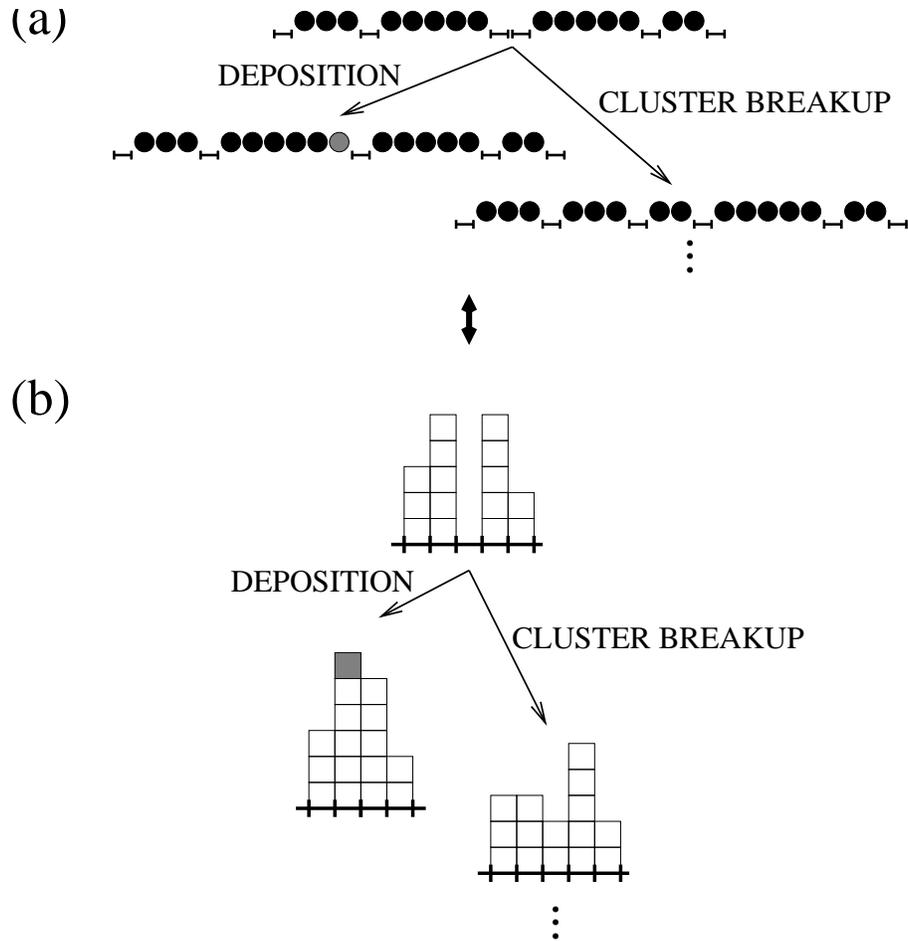}
\caption{(a) Possible events after the formation of a double vacancy
in the original particle picture: 
deposition of a new particle (grey circle) and one of the possible events
of cluster breakup (left cluster) are illustrated. (b) The corresponding
events in the column picture. Notice that deposition of a new particle
(grey square) leads to the decrease of the lattice length in this picture.}
\label{fig7}                        
\end{center}
\end{figure}

\begin{figure}
\epsfxsize=12cm
\begin{center}
\leavevmode
\epsffile{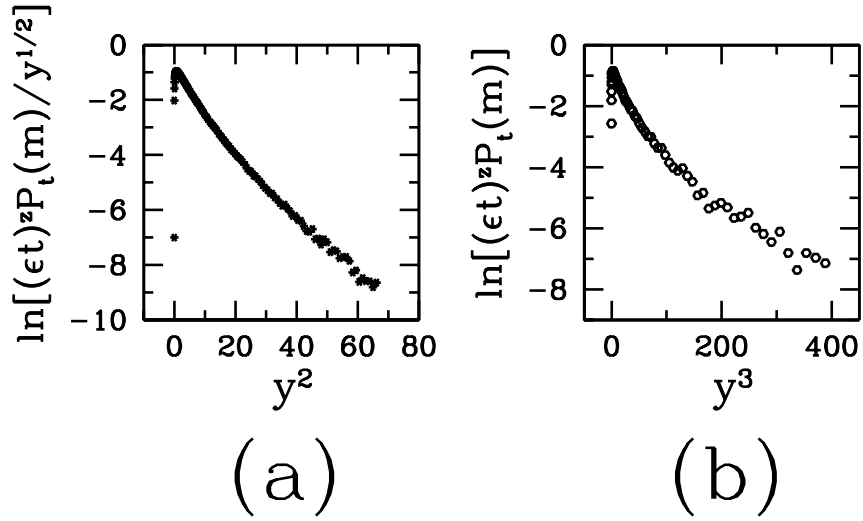}
\caption{Scaled probability of a cluster of size $m$ as a function of the
scaled mass according to the prediction of the independent interval
approximation (Eqs. \ref{disttheory} and \ref{exptheory}), for:
(a) $\alpha=2$, $\epsilon =0.001$, $z=1/4$, $t=4\times{10}^7$;
(b) $\alpha=4$, $\epsilon =0.001$, $z=1/6$, $t=5\times{10}^8$.}
\label{fig8}                        
\end{center}
\end{figure}

\begin{figure}
\epsfxsize=12cm
\begin{center}
\leavevmode
\epsffile{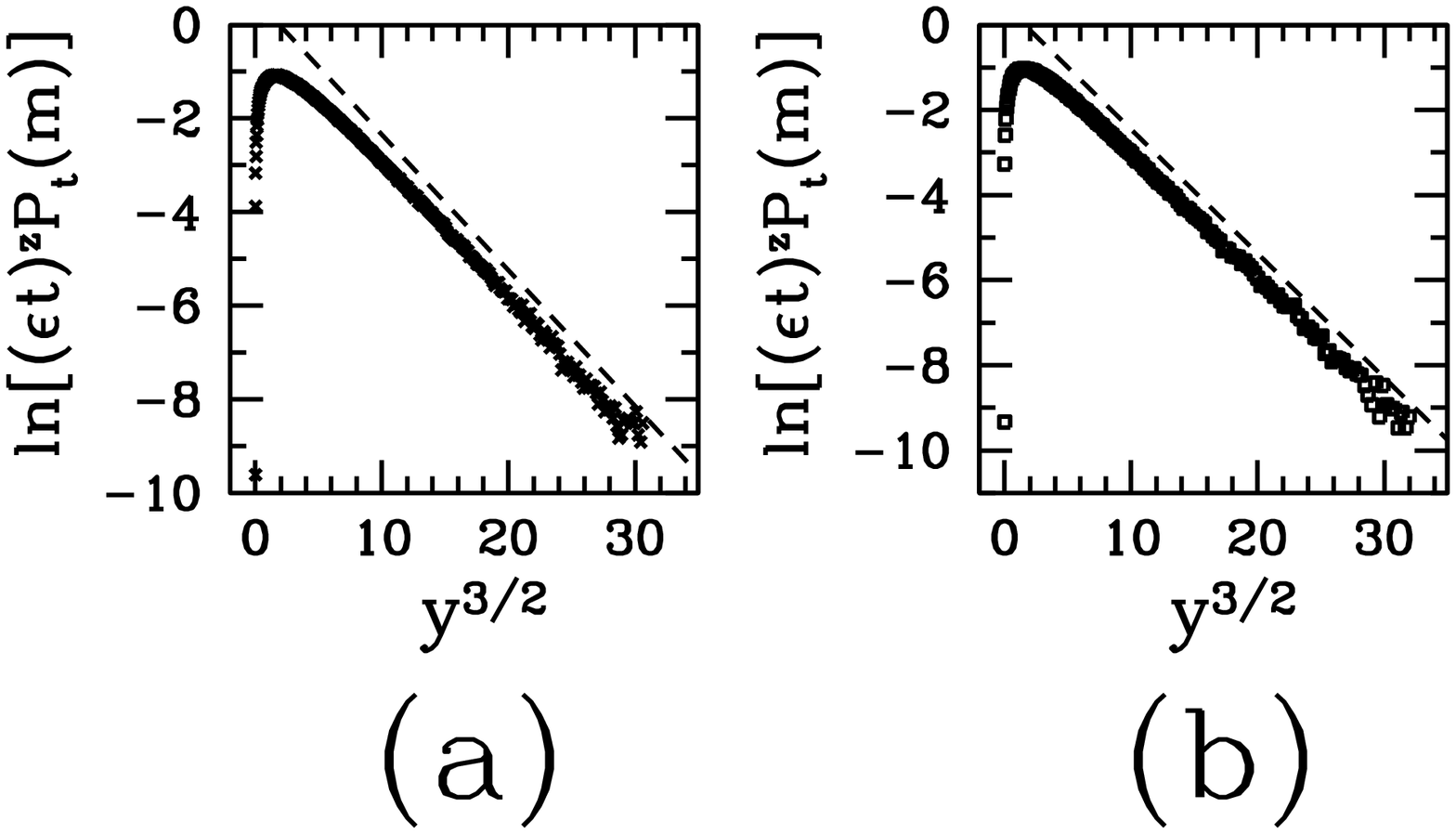}
\caption{Scaled probability of a cluster of size $m$ as a function of 
the scaled mass $y^{3/2}$, for:
(a) $\alpha=0.5$, $\epsilon =0.001$, $z=0.4$, $t=4\times{10}^6$;
(b) $\alpha=1$, $\epsilon =0.001$, $z=1/3$, $t=4\times{10}^6$.}
\label{fig9}                        
\end{center}
\end{figure}

\begin{figure}
\epsfxsize=12cm
\begin{center}
\leavevmode
\epsffile{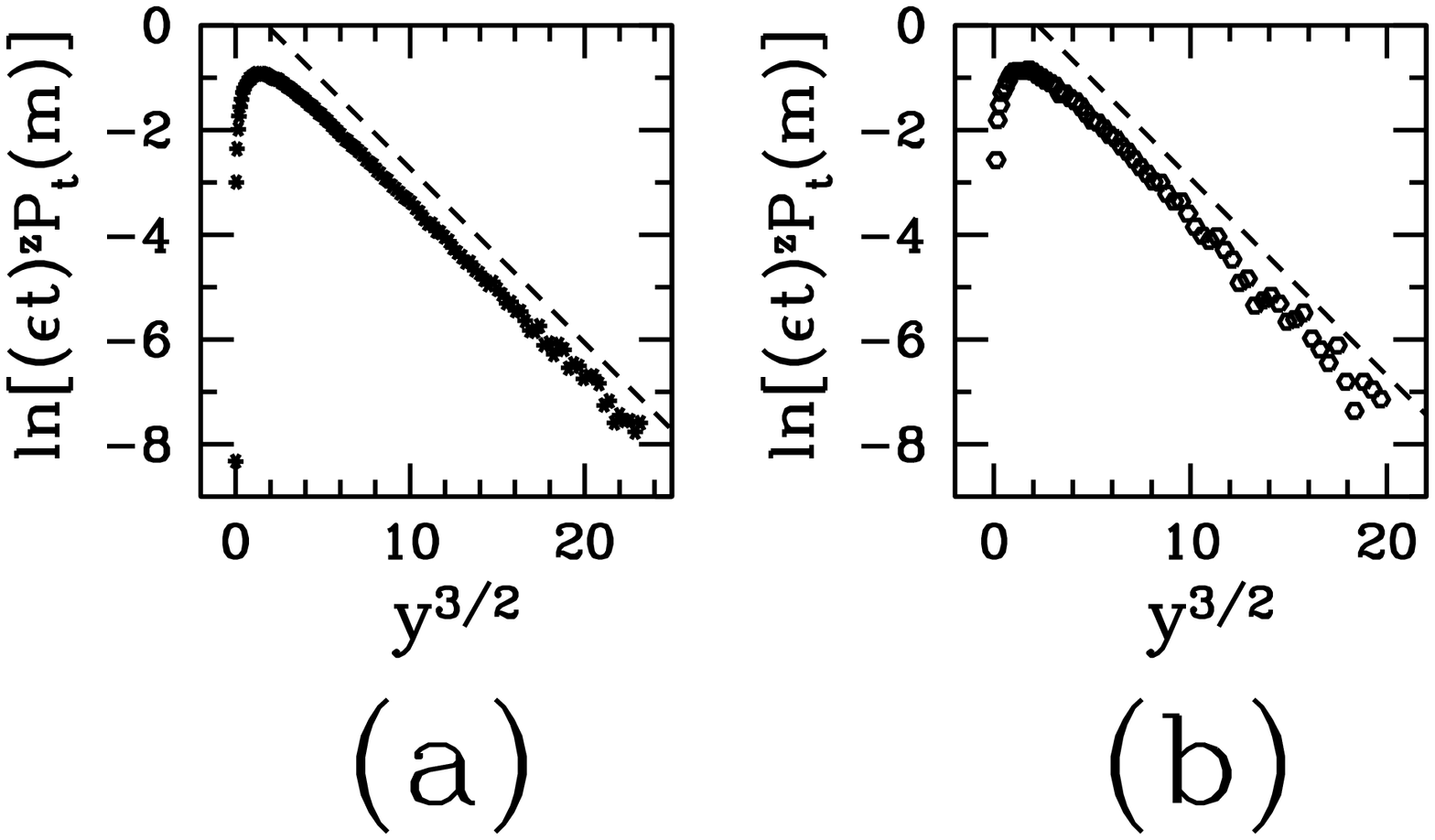}
\caption{Scaled probability of a cluster of size $m$ as a function of 
the scaled mass $y^{3/2}$, for:
(a) $\alpha=2$, $\epsilon =0.001$, $z=1/4$, $t=4\times{10}^7$;
(b) $\alpha=4$, $\epsilon =0.001$, $z=1/6$, $t=5\times{10}^8$.}
\label{fig10}                        
\end{center}
\end{figure}


\begin{references}

\bibitem{robin}
R. B. Stinchcombe, Adv. Phys. {\bf 50}, 431 (2001).

\bibitem{ritort}
F. Ritort and P. Sollich, Adv. Phys. {\bf 52}, 219 (2003).

\bibitem{mevansreview}
M. R. Evans, J.Phys:Condens. Matter {\bf 14} 1397 (2002).

\bibitem{dhar}
D. Dhar, Physica A {\bf 315} 5 (2002).

\bibitem{koponen}
I. Koponen, M. Rusanen, and J. Heinonen, Phys. Rev. E {\bf 58}, 4037 (1998).

\bibitem{coarsen1}
F. D. A. Aar\~ao Reis and R. B. Stinchcombe, Phys. Rev. E {\bf 70}, 036109
(2004).

\bibitem{cv}
S. Clarke and D. D. Vvedensky, J. Appl. Phys. {\bf 63}, 2272 (1988).

\bibitem{family}
F. Family, P. Meakin, and J. M. Deutch, Phys. Rev. Lett. {\bf 57}, 727 (1986).

\bibitem{meakin}
P. Meakin and M. H. Ernst, Phys. Rev. Lett. {\bf 60}, 2503 (1988).

\bibitem{leyvraz}
F. Leyvraz and S. Redner, Phys. Rev. Lett. {\bf 88}, 068301 (2002).

\bibitem{bennaim}
E. Ben-Naim and P. L. Krapivsky, Phys. Rev. E {\bf 68}, 031104 (2003).

\bibitem{bramson}
M. Bramson and J. L. Lebowitz, Phys. Rev. Lett. {\bf 61}, 2397 (1988);
{\bf 62}, 694 (1989); J. Stat. Phys. {\bf 62}, 297 (1991).

\bibitem{bray1}
A. J. Bray and R. A. Blythe, Phys. Rev. Lett. {\bf 89}, 150601 (2002).

\bibitem{bray2}
 R. A. Blythe and A. J. Bray, Phys. Rev. E {\bf 67}, 041101 (2003).

\bibitem{lai}
Z. W. Lai, G. F. Mazenko, and O. T. Valls, Phys. Rev. B {\bf 37}, 9481
(1988).

\bibitem{shore}
J. D. Shore, M. Holzer, and J. P. Sethna, Phys. Rev. B {\bf 46}, 11376 (1992).

\end{references}
\end{document}